\RequirePackage[l2tabu, orthodox]{nag}
\RequirePackage{snapshot}

\documentclass[9pt,onecolumn]{extarticle}

\makeatletter
\if@twocolumn
  \usepackage[dvips,letterpaper,top=0.5in, bottom=0.5in, left=0.75in, right=0.5in,includefoot,heightrounded]{geometry}
\else
  \usepackage[dvips,letterpaper,margin=1in,includefoot,heightrounded]{geometry}
\fi

\usepackage{srcltx}

\usepackage[russian,portuges,english]{babel}

\iflanguage{portuges}
    {\newcommand{\keywordname}{Palavras-chaves}}
    {\newcommand{\keywordname}{Keywords}}

\usepackage{amsmath}
\usepackage{amssymb,amsfonts}

\usepackage{abstract}

\usepackage{graphicx}
\usepackage[usenames,dvipsnames,svgnames,x11names]{xcolor}
\usepackage{subfigure}

\usepackage{booktabs}

\usepackage{setspace}
\usepackage{flushend}

\usepackage{cite}

\usepackage{hyperref}
\usepackage[normalem]{ulem}
\usepackage{bm}

\usepackage{enumerate}

\usepackage{multirow}
\usepackage[noend]{algpseudocode}

\usepackage{listings}

\usepackage[short,12hr]{datetime}
       \usepackage{fouriernc}

\newcommand{\E}{\operatorname{E}}
\newcommand{\var}{\operatorname{Var}}
\newcommand{\diag}{\operatorname{diag}}

\makeatletter

\newcommand{\printtitle}{%
\makeatletter
\if@twocolumn

\twocolumn[%
  \maketitle
  \begin{onecolabstract}
    \myabstract
  \end{onecolabstract}
  \begin{center}
    \small
    \textbf{\keywordname}
    \\\medskip
    \mykeywords
  \end{center}
  \bigskip
]
\saythanks
\else
  \maketitle
  \begin{onecolabstract}
    \myabstract
  \end{onecolabstract}
  \begin{center}
    \small
    \textbf{\keywordname}
    \\\medskip
    \mykeywords
  \end{center}
  \bigskip
  \onehalfspacing
\fi
\makeatother
}

\author{%
B. G. Palm%
\thanks{Programa de P\'os-gradua\c{c}\~ao em Estat\'istica,
Universidade Federal Pernambuco
and
Department of
Telecommunications,
Aeronautics Institute
of Technology (ITA), Brazil
(E-mail: \protect\url{brunagpalm@gmail.com}).}
\and
F. M. Bayer%
\thanks{Departamento de Estat\'istica
and LACESM,
Universidade Federal de Santa Maria, Brazil
(E-mail: \protect\url{bayer@ufsm.br}).}
\and
R. J. Cintra%
\thanks{Signal Processing Group,
Departamento de Estat\'istica,
Universidade Federal Pernambuco, Brazil
(E-mail: \protect\url{rjdsc@de.ufpe.br}).}
}

\title{%
Improved Point Estimation for the Rayleigh Regression Model}

\newcommand{\myabstract}{%
The Rayleigh regression model was recently proposed for modeling
amplitude values of synthetic aperture
radar~(SAR) image pixels.
However,
inferences from such model
are based on the maximum likelihood estimators,
which can be biased for small signal lengths.
The Rayleigh
regression model for SAR images
often takes into account small pixel windows,
which may lead to inaccurate results.
In this letter, we introduce
bias-adjusted estimators tailored for the
Rayleigh regression model
based on:
(i)~the Cox and Snell's method; (ii)~the Firth's
scheme; and (iii)~the parametric bootstrap method.
We present
numerical experiments considering synthetic
and actual SAR
data sets.
The bias-adjusted estimators
yield nearly unbiased estimates and accurate modeling
results.
}

\newcommand{\mykeywords}{%
Bias correction,
Rayleigh regression model,
SAR images,
Small signal lengths inferences
}

\date{}

\begin{document}

\printtitle

\section{Introduction}

The classical
linear regression model
is widely employed
to estimate
an
unknown
and
deterministic
parameter vector
assuming
the Gaussian distribution~\cite{wiesel2008}.
However,
practical contexts
often exhibit non-Gaussian behavior.
An alternative to the Gaussian model
is provided by the Rayleigh distribution
which is capable of
characterizing
asymmetric,
continuous,
and
nonnegative
signals,
such as
the
amplitude
values of
synthetic aperture radar~(SAR) image
pixels~\cite{gomes2018,
sumaiya2018}.
The Rayleigh regression model was proposed in~\cite{Palm2019},
where a methodology for point estimation,
large data record results,
and
goodness-of-fit measures
were presented
and discussed
in the context of SAR image detection.

Parameter inference based on
the
Rayleigh regression model
can be achieved by means
of the
maximum likelihood estimation,
inheriting its
good asymptotic properties
for large signal lengths.
However,
if the signal length~$N$
is small, then the
maximum likelihood estimators~(MLE)
present
a
bias in the order of~$N^{-1}$,
which can be regarded
as
problematic~\cite{CordeiroCribari2014}.
For instance,
in~\cite{Palm2019},
the detection
of different land cover types
in  SAR images
was performed based on
pixel windows
of
more than~$126$
pixels.
However,
if smaller windows,
such as
$3 \times 3$,
are selected for
Rayleigh regression model
parameter estimation,
then
the obtained
results
can be severely
biased.

An approach
to address this issue
is by means of inferential
corrections~\cite{CordeiroCribari2014}.
Three
widely
bias-adjusted
methods
are
the Cox and Snell's~\cite{cox1968},
Firth's~\cite{Firth1993},
and parametric bootstrap~\cite{Efron1979}
schemes.
The Cox and Snell's method
is an analytical approach
used to obtain
second order
corrected estimators.
The Firth's method
is a preventive method~\cite{Firth1993}
of bias reduction
which modify the score function
before obtaining parameter estimates
based on the analytical second order biases
of the~MLE~\cite{Firth1993}.
Finally,
the bootstrap method is
a computationally intensive
method based on resampling,
being suitable for
inferential corrections
when~$N$ is small~\cite{Efron1979}.

To the best of
our knowledge,
the literature lacks
bias-adjusted estimators
for the parameters of
the Rayleigh regression model.
In this paper,
our chief goal is to obtain
accurate point
estimation approaches
to address this
literature gap.

\section{The Rayleigh Regression Model}
\label{s:rrmodel}

The Rayleigh regression model
was proposed in~\cite{Palm2019}
and can be defined as follows.
Let~$Y$
be
a
Rayleigh distributed random variable
with mean parameter~$\mu > 0$.
The
probability density function
of the mean-based Rayleigh distribution
is given by~\cite{Palm2019}
\begin{align}
\label{e:den}
f_Y(y;\mu )
=
\frac{\pi y }{2 \mu  ^2}
\exp\left(-\frac{\pi y ^2}{4 \mu ^2}\right)
,
\end{align}
where~$y > 0$
is the observed signal value.
The mean and variance
of~$Y$
are given,
respectively,
by
$\E(Y) = \mu
$
and
$\var(Y) = \mu ^2 \left(\frac{4}{\pi}-1 \right)
$.

Let~$Y[1], Y[2], \ldots, Y[N]$
be
independent random variables,
where each~$Y[n]$
assumes values~$y[n]$
and
follows
the
Rayleigh density in~\eqref{e:den}
with mean~$\mu[n]$,
$n=1,2,\ldots,N$.
The Rayleigh regression model is
defined
assuming that the mean of
the observed output signal~$Y[n]$
can be written as
\begin{align*}
\eta[n]
=
g(\mu[n]) =
\sum_{i=1}^{k}  \beta_i  x_{i}[n]
,
\quad
n=1,2,\ldots, N,
\end{align*}
where~$k<N$ is the
number
of covariates considered
in the model,
$\bm{\beta} = (\beta_1, \beta_2, \ldots, \beta_k)^{\top}$
is the
vector
of unknown
linear parameters,
$\mathbf{x}[n]=(x_{1}[n], x_{2}[n], \ldots, x_{k}[n])^\top$
is the
vector
of
independent
input
variables,~$g:\mathbb{R}^+\!\rightarrow\mathbb{R} $
is a strictly monotonic and twice
differentiable link function,
and~$\eta[n]$
is the
linear predictor~\cite{Palm2019}.

Parameter estimation
can be performed using
the maximum likelihood method,
as discussed in~\cite{Palm2019}.
The estimated vector~$\widehat{\bm{\beta}}$
is obtained by maximizing
the logarithm of the likelihood function.
The log-likelihood function
of the parameter vector~$\bm{\beta}$
for the observed signal is
$
\ell(\bm{\beta})=\sum_{n=1}^{N} \ell_n (\mu[n])
$,
where
$
\ell_n (\mu[n])
=
\log\left(\frac{\pi}{2}\right) +
\log(y[n])
- \log(\mu[n]^2)-\frac{\pi y[n]^2}{4 \mu[n]^2}
$.
The score vector can be written as
$
U(\bm{\beta})
=
\mathbf{X}^\top
\cdot
\mathbf{T}
\cdot
\mathbf{v}
,
$
where~$\mathbf{X}$ is the~$N \times k$
matrix whose~$n$th row
is~$\mathbf{x}[n]^\top$,
$
\mathbf{T}
=
\diag\left\{
\frac{1}{g'(\mu[1])},
\frac{1}{g'(\mu[2])},
\ldots,
\frac{1}{g'(\mu[N])}
\right\}
$,
and~$\mathbf{v}
=
\left(\frac{\pi y[1]^2}{2 \mu[1]^3}-\frac{2}{\mu[1]},
\frac{\pi y[2]^2}{2 \mu[2]^3}-\frac{2}{\mu[2]},
\ldots,
\frac{\pi y[N]^2}{2 \mu[N]^3}-\frac{2}{\mu[N]} \right)^\top$.
Finally,
the Fisher information matrix is given by
$
\mathbf{I}(\bm{\beta})=
\mathbf{X}^\top
\cdot
\mathbf{W}
\cdot
\mathbf{X}
$,
where
$\mathbf{W}
=
\diag\left\{
\frac{4}{\mu[1]^2}\left(\dfrac{\operatorname{ d }
\mu[1]}{\operatorname{ d }\eta[1]}\right)^2,
\ldots,
\frac{4}{\mu[N]^2}\left(\dfrac{\operatorname{ d }
\mu[N]}{\operatorname{ d }\eta[N]}\right)^2
\right\}$.
Further mathematical properties,
including  large data record results
and optimization procedures,
are detailed in~\cite{Palm2019}.

\section{Bias Correction of MLE}
\label{s:bias}

Generally,
for small~$N$,
the~MLE
$\bm{\widehat{\beta}}$
for
$\bm{\beta}$
may be biased
and the associated bias
can be expressed
as
$
B(\widehat{\bm{\beta}})
=
\E[\widehat{\bm{\beta}}]-\bm{\beta}
$~\cite{CordeiroCribari2014}.
The
Cox and Snell's
bias correction
formula
for the~$a$th component
of~$\bm{\widehat{\beta}}$
is given by
\begin{align}
\label{e:vies}
B(\widehat{\beta}_a)
=
\sum \limits _{r,s,u}
\kappa ^{ar}
\kappa ^{su}
\left\lbrace
\kappa_ {rs} ^ {(u)}
-
\frac{1}{2}
\kappa_{rsu}
\right\rbrace
,
\end{align}
where
$\kappa_ {rs}
=
\E
\left(
\frac{\partial ^2 \ell}{\partial \beta_r \partial \beta_s}
\right)
$,
$\kappa_ {rs}^ {(u)}
=
\frac{\partial \kappa_ {rs}}{\partial \beta_u}
$,
$\kappa_ {rsu}
=
\E
\left(
\frac{\partial ^3 \ell}{\partial \beta_r \partial \beta_s
\partial \beta_u}
\right)
$,
$-\kappa ^{ar}$ and $-\kappa ^{su}$
are
the~$(a,r)$
and~$(s,u)$
elements of the inverse of
the Fisher information
matrix,
respectively~\cite{cox1968}.
The cumulants
obtained for the
Rayleigh regression model
can be found in the Appendix.

The second order
bias of~$\widehat{\bm{\beta}}$
is defined
as
\begin{align*}
B(\widehat{\bm{\beta}})
=
\mathbf{I}^{-1} (\widehat{\bm{\beta}})
\cdot
\mathbf{X}^\top
\cdot
\mathbf{W}
\cdot
\delta
,
\end{align*}
where~$\delta$
is
the main diagonal
of~$
\mathbf{X}
\cdot
\mathbf{I}^{-1} (\bm{\beta})
\cdot
\mathbf{X}^\top
$
and
$
\mathbf{W}
=
\diag
\left[
-
\frac{2}{\mu[n]^3}
\left(
\frac{\operatorname{ d } \mu [n]}{\operatorname{ d } \eta [n]}
\right)^3
-
\frac{2}{\mu [n] ^2}
\left(
\frac{\operatorname{ d } \mu [n]}{\operatorname{ d } \eta [n]}
\right)^2
\cdot
\frac{\partial}{\partial \mu [n]}
\left(
\frac{\operatorname{ d } \mu [n]}{\operatorname{ d } \eta [n]}
\right)
\right ]
$.
Replacing the
unknown parameters
by their
MLE,
we have the~MLE~$\widehat{B}(\widehat{\bm{\beta}})$
for~$B(\widehat{\bm{\beta}})$,
and
bias-adjusted estimators
can be derived
removing~$\widehat{B}(\widehat{\bm{\beta}})$
of~$\widehat{\bm{\beta}}$~\cite{CordeiroCribari2014}.
Hence,
corrected estimators
based on the Cox and Snell's method,~$\widetilde{\bm{\beta}}$,
are obtained
as
$
\widetilde{\bm{\beta}}
=
\widehat{\bm{\beta}}
-
\widehat{B}(\widehat{\bm{\beta}})
$~\cite{cox1968}.
The Firth's method
removes the
second-order bias
by
modifying the original
score function~$U(\bm{\beta})$
according to
$
U^\ast(\bm{\beta})
=
U(\bm{\beta})
-
\mathbf{I} (\bm{\beta})
\cdot
B ( \bm{\widehat{\beta}})
$~\cite{Firth1993},
where~$B ( \bm{\widehat{\beta}})$
is the
second-order bias
computed in~\eqref{e:vies}.
The roots
of the modified score function~$U^\ast(\bm{\beta})$
constitute the corrected estimator~$\widehat{\bm{\beta}}^\star$
according to the Firth's method.
In the bootstrap bias correction method,
the bias estimation~$\widehat{B}(\widehat{\bm{\beta}})$
is numerically obtained through Monte Carlo simulations.
A bootstrap estimate
of the bias can be obtained by
$
\widehat{B}_{\text{boot}}(\widehat{\bm{\beta}})
=
\bar{\bm{\beta}}^{\ast} -\widehat{\bm{\beta}}
$,
where $\bar{\bm{\beta}}^{\ast}=\frac{1}{R}
\sum \limits _{b=1}^{R}
\widehat{\bm{\beta}}_b$,~$R$ is the number of bootstrap replications,
and~$\widehat{\bm{\beta}}_b$
is the estimated values of~$\bm{\beta}$
in each bootstrap replication.
Thus,
the corrected
estimator
based on the bootstrap
method is given by
$
\widehat{{\bm{\beta}}}^\ast
=
\widehat{\bm{\beta}}
-
\widehat{B}_{\text{boot}}
(\widehat{\bm{\beta}})
=
\widehat{\bm{\beta}}
-
(\bar{\bm{\beta}}^{\ast}
-
\widehat{\bm{\beta}})
=
2
\widehat{\bm{\beta}}
-
\bar{\bm{\beta}}^\ast
$~\cite{Efron1979}.
The above described bias-adjusted estimators
share the same asymptotic
properties
with the usual~MLE
but are less biased for small~$N$~\cite{Efron1994}.

\section{Numerical Experiments}
\label{s:simulations}

\subsection{Synthetic Data Modeling}

Monte Carlo simulations
were employed
to evaluate the
original MLE
performance
of the Rayleigh regression model
parameters
and
their bias-adjusted versions.
Synthetic SAR data was generated under two scenarios.
Each scenario
aimed
at capturing
asymmetric distributions.
For
Scenario~1,
we selected the following parameters
values:
$\beta_1= 0.5$,~$\beta _2 =0.5 $,~$\beta_3 = 1$,
and covariates generated from
the binomial distribution;
whereas,
for
Scenario~2,
we adopted~$\beta_1= 2.5$,~$\beta _2 =1.5 $,
and
covariates generated from
the Rayleigh distribution.
We have
for Scenarios~1 and~2,
skewness
about~$4$
and~$3$, respectively.
The covariates
values
were
kept constant
for all Monte Carlo replications
and
the log link function
was employed.
The number of Monte Carlo and bootstrap replications
were set equal to~$5{,}000$ and~$1{,}000$,
respectively,
and the
signal
lengths
considered
were~$N \in \{
9; 25; 49
\}$.
Such blocklengths
are popular choices
of window sizes
in SAR image processing~\cite{park1999,li2012,dimas2019}.

The percentage relative bias (RB\%)
and the root mean squared error (RMSE)
were adopted
as figures of merit to numerically evaluate
the point estimators.
Table~\ref{t:t1}
presents the simulation results
for point estimation of the
Rayleigh regression model
parameters
for Scenarios~1 and~2.
We notice
that the MLE
can be strongly
biased
for small~$N$.
For instance,
for Scenario~1
and~$N=9$,
the relative
biases
of the MLE
are approximately
(in absolute values)~$10\%$,~$24\%$, and~$1\%$
for~$\widehat{\beta}_1$,~$\widehat{\beta}_2$,
and~$\widehat{\beta}_3$,
respectively.
For this same signal length,
the relative bias of~$\widetilde{\beta}_2$,~$\widehat{\beta}_2^\star$,
and~$\widehat{\beta}_2^\ast$
are (in absolute values)~$9\%$,~$9\%$,
and~$7\%$,
respectively.
In general,
the bias-corrected estimators
present values closer to the true parameters
when compared to
the MLE
and
have
similar performance
in terms of relative bias
and RMSE.

\begin{table*}
	\setlength{\tabcolsep}{2.8pt}
	\centering
\caption{
	Results of the Monte Carlo
	simulation for point estimation
	of
	Scenarios~1 and~2.
	Best results are highlighted}
\label{t:t1}
\begin{tabular}{cccccc|cccc|cccc}
	\toprule
	&	& $\widehat{\beta}_1$  & $\widetilde{\beta}_1$ &
	$\widehat{\beta}_1^\star$
	& $\widehat{\beta}_1^\ast$
	& $\widehat{\beta}_2$ & $\widetilde{\beta}_2$ &
	$\widehat{\beta}_2^\star$
	& $\widehat{\beta}_2^\ast$
	& $\widehat{\beta}_3$ &
	$\widetilde{\beta}_3$ &
	$\widehat{\beta}_3^\star$
	& $\widehat{\beta}_3^\ast$
	\\
	\midrule
	\multicolumn{14}{c}{$N=9$} \\
	\midrule
	\multirow{2}{*}{Scenario 1}&	RB$(\%)$  & $-10.0546$ &    $0.9665$ &
	$\mathbf{-0.4822}$ &   $ 1.0152$ &  $-24.3255$ &   $-8.9817$ &   $-8.8812$
	& $\mathbf{-7.1089}$ &  $ -1.4237$ &  $\mathbf{1.0249}$ & $2.5234$ &  $1.2595$   \\
	&	RMSE  &   $0.7083$ &    $0.7049 $ &   $0.7068$ &    $0.7051$ &
	$0.7960$ &    $0.7927$ &    $0.7924$ &    $0.7922$ &    $0.8083 $ &
	$0.8055$ & $0.8087$ &    $0.8062$  \\
&	RB$(\%)$  &
	$-1.6082$  & $-0.0595$  & $-0.1941$ &  $\mathbf{-0.0252}$ &  $-2.4411$
	& $-0.5491$ &   $\mathbf{0.2736}$  & $-0.4622$ & -- & -- & -- & --\\
	\multirow{-2}{*}{
	Scenario 2} 	& RMSE &
	$0.4625$  &  $0.4599$    &$0.4610$  &  $0.4600$    &$0.6979$
	&   $0.6959$  &  $0.6980$&    $0.6961$ & -- & -- & -- & --\\
	\midrule
	\multicolumn{14}{c}{$N=25$} \\
	\midrule
	\multirow{2}{*}{Scenario 1}&		RB$(\%)$  &    $-1.2113$ &   $ 1.1810$ &
	$\mathbf{0.6506}$  &   $ 1.4531$ &  $-13.9903$ &   $-4.1510$
	&   $\mathbf{-2.1974}$ &   $-4.5960$        &  $-1.8112$
	&   $-0.3642$ & $\mathbf{-0.0567}$ & $0.4670$ \\
	&		RMSE & $0.3319$ &    $0.3317$ &    $0.3317$ &
	$0.3316$ &   $ 0.3568 $ &   $0.3562$ &    $0.3563$ &    $0.3565$ &
	$0.3554$ &    $0.3548$ & $0.3548$ &    $0.3549$ \\
& 	RB$(\%)$  &
	$-0.2640$  &  $0.1544$ &   $\mathbf{0.1150} $ &  $0.1812$  & $-1.3497$ &
	$-0.3138$ &  $\mathbf{-0.1566}$  & $-0.3695$ & -- & -- & -- & --\\
			\multirow{-2}{*}{
	Scenario 2}	&	  RMSE &
	$0.2319$   & $0.2318$  &  $0.2319$  & $ 0.2319$ &   $0.3439$  &
	$0.3432$  &  $0.3434$   & $0.3432$ & -- & -- & -- & --\\
	\midrule
	\multicolumn{14}{c}{$N=49$} \\
	\midrule
	\multirow{2}{*}{Scenario 1}&
	RB$(\%)$  & $-0.9191$ &   $\mathbf{-0.0037}$ &
	$-0.1914$ &   $ 0.0997$ &   $-7.0768$ &  $ -1.3383$ &   $\mathbf{-0.6904}$  &
	$ -1.5395$ &   $-0.6448$ &    $0.2236$ & $0.3203$ &    $\mathbf{0.1924}$  \\
	&		RMSE &  $0.2226$ &    $0.2225$ &    $0.2225$ &    $0.2226$ &
	$0.2381$ &    $0.2380$ &    $0.2380$ &    $0.2381$ &    $0.2365$ &    $0.2362$ &
	$0.2362$ &   $ 0.2364$  \\
& 	RB$(\%)$  &
	$-0.1297$  &  $0.0580$  &  $\mathbf{0.0462}$  &  $0.0716$ &  $-0.7194$
	& $-0.1266$ &  $\mathbf{-0.0821}$  & $-0.1552$ & -- & -- & -- & --\\
	\multirow{-2}{*}{
	Scenario 2}	&	RMSE &
	$0.1588$  &  $0.1588$  &  $0.1588$  &  $0.1589$  &  $0.2306$
	&  $0.2304$ &   $0.2304$  &  $0.2305$ & -- & -- & -- & --\\
	\bottomrule
\end{tabular}
\end{table*}

To evaluate the
overall
performances
of the four estimators
for each value of~$N$,
we employed the
integrated relative bias squared
norm~(IRBSN) figure of merit~\cite{cribari2002},
which is defined as~$
\text{IRBSN }
=
\sqrt{\frac{1}{k}\sum_{i=1}^k
	\text{RB}(\widehat{\beta}_i)^2}
$,
where~$\text{RB}
(\widehat{\beta}_i)$,~$i = 1,2, \ldots , k$,
corresponds
to the
values of RB\% of each
estimator.
The values of~IRBSN for Scenarios~1 and~2
are given in~Table~\ref{t:t2}.
The corrected estimators
excel in terms
of IRBSN.
Additionally, among the evaluated estimators,
the ones obtained by the
Firth's
method present the smallest values of
IRBSN in five of the six evaluated scenarios.

\begin{table}
	\centering
		\setlength{\tabcolsep}{2.8pt}
	\caption{
		Integrated relative bias squared norm
		results.
		Best results are
		highlighted
	}
	\label{t:t2}
	\begin{tabular}{cccccc}
		\toprule
	&	& MLE &
		Cox and Snell &
		Firth &
		Bootstrap  \\
		\midrule
	&	$N=9$ &  $15.2190$ &    $5.2490$ &    $5.3378$ &    $\mathbf{2.8764}$\\
Scenario~1 &		$N=25$ & $8.1747$ &    $2.5005$ &    $\mathbf{1.3235}$ &    $2.7960$ \\
&		$N=49$ &  $4.1369$ &    $0.7834$ &    $\mathbf{0.4531}$ &    $1.0368$ \\
		\midrule
&		$N=9$ & $2.0670$   & $0.3905$  &  $\mathbf{0.2372}$  &  $0.2910$ \\
	Scenario~2 &	$N=25$ &  $0.9725$ &    $0.2473$ &   $ \mathbf{0.1374}$ &   $ 0.2910$ \\
	&	$N=49$ &  $ 0.5169$  &  $0.0985 $  & $\mathbf{0.0666}$ &   $ 0.1209$\\
		\bottomrule
	\end{tabular}
\end{table}

\subsection{SAR Image Modeling}
\label{s:sar}

In this section,
we present
image modeling experiments considering
two actual SAR
data sets.
For such,
we considered
(i)~the
San Francisco Bay~(SF)
associated
to
horizontal~(HH),
vertical~(VV),
and~HV
polarization channels
SAR images;
and
(ii)~an
image
obtained from
CARABAS~II,
a
Swedish ultrawideband~(UWB)
very-high frequency~(VHF) SAR system.
More details about these data sets
can be found in~\cite{Cintra2013,Lundberg2006}.
As the correlation between different channels
has been considered
as a scheme
for
detection of man-made targets
and land cover classification
in SAR imagery~\cite{xiang2016built},
we used
the HV channel
to describe the
mean of the
amplitude values
of HH and VV channels---the resulted models
are referred to as HH~$\sim$~HV
and
VV~$\sim$~HV, respectively.
Additionally,
San Francisco HV associated image is well correlated
with the
HH and VV
channels:~$0.8268$ and~$0.7593$,
respectively.
To specify the mean of the amplitude
values of the CARABAS~II
SAR image
associated
to mission one and pass one,
we fitted the regression model
considering
as a covariate,
an
image
with the same fight pass
of the evaluated CARABAS~II
image~(mission two and pass one)---the correlation value between
both images is $ 0.4719$.
As the Rayleigh distribution
is suitable for representing homogeneous areas~\cite{oliver2004},
such as the sea/water ground type~\cite{Cintra2013},
we selected windows
of~$3 \times 3$ and~$7 \times 7$ pixels
related to the water area
to fit the models.

The adjusted model results can be found in
Table~\ref{t:fit}.
We notice that the MLE
and Firth-based estimators
present different estimate values
in all evaluated scenarios.
For instance,
considering the window
of~$7\times 7$ pixels
in the $\text{HH} \sim \text{HV}$ model,
$\widehat{\beta}^\star_2$
shows a value
about~$1820\%$
bigger than~$\widehat{\beta}_2$.
To quantify the performance of considered fitted models,
we employed the
estimated
root mean square error
measure
computed between~$y[n]$
and~$\widehat{\mu}[n]$---here referred to as~$\widehat{\text{RMSE}}$.
For comparison purposes,
we also fitted the
standard Gaussian regression model---widely employed
in signal modeling---and the Rayleigh distribution.
The
$\widehat{\text{RMSE}}$
values
for the considered models are given in
Table~\ref{t:fit2};
for all evaluated models,
the Firth-based estimators
present smaller values of~$\widehat{\text{RMSE}}$.

\begin{figure}
	\centering
	\subfigure[CARABAS~II]{\includegraphics[scale=0.223]{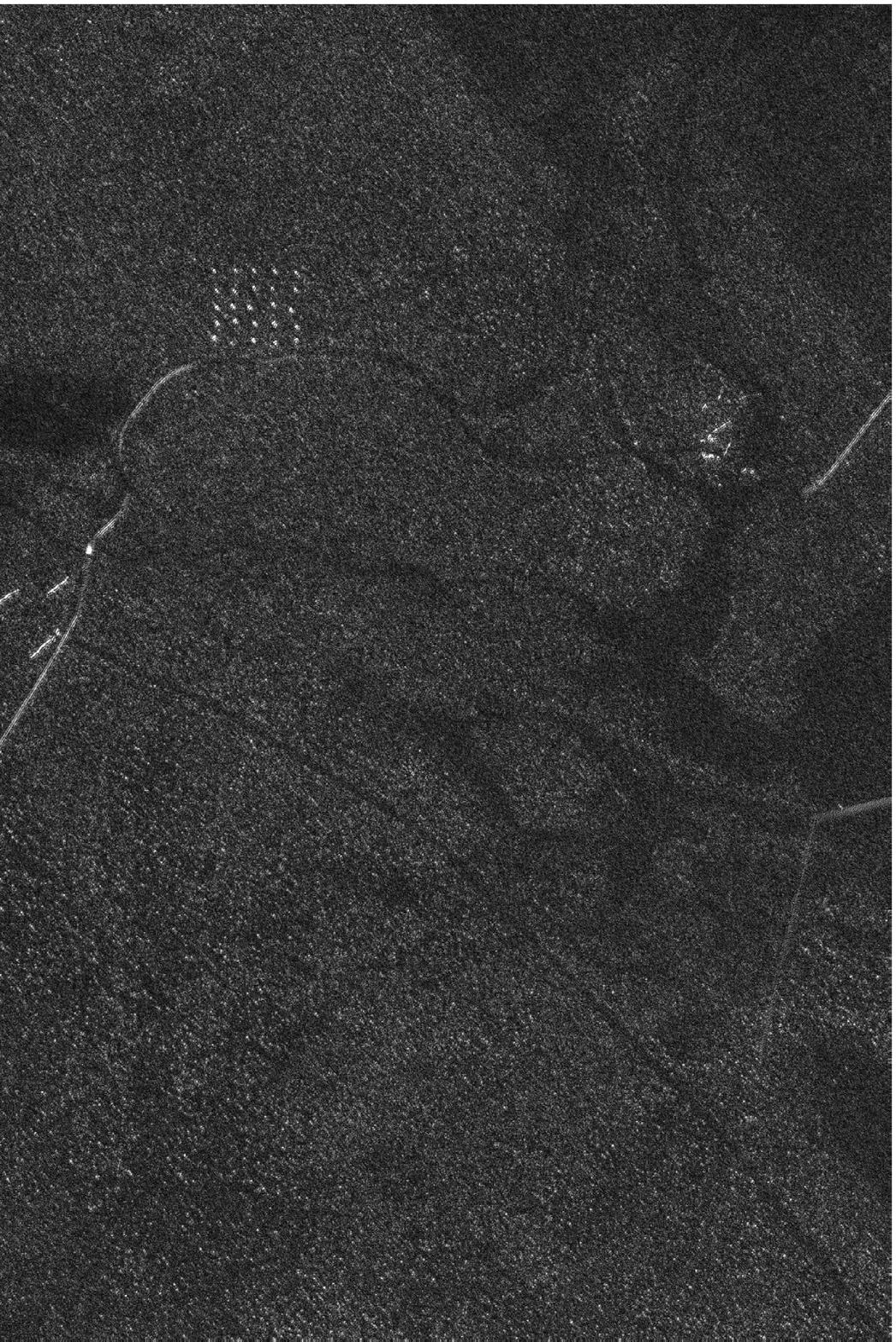}}
	\subfigure[San Francisco]{\includegraphics[scale=0.13]{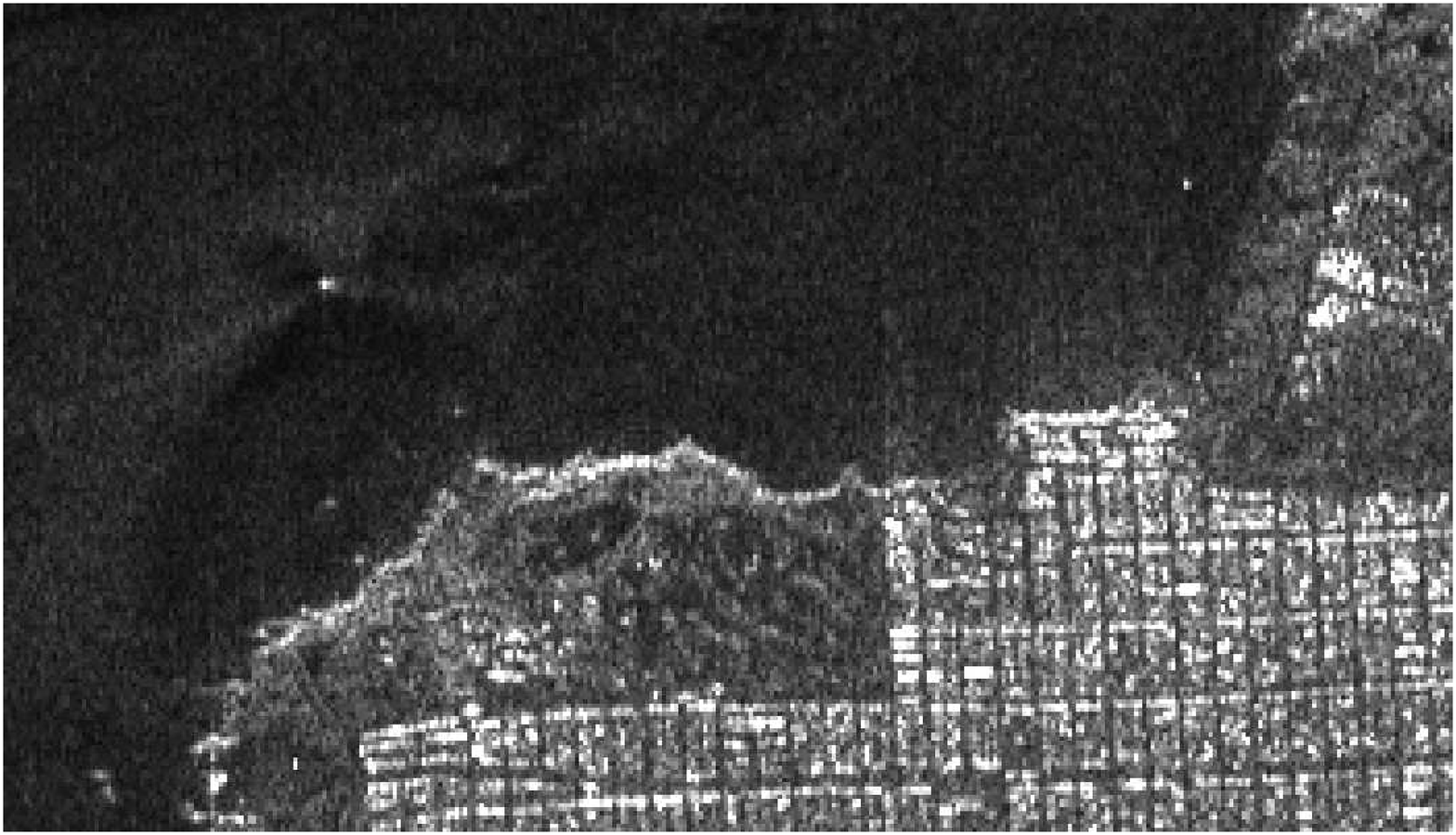}}
	\caption{San Francisco and CARABAS~II SAR images associated to
		HH polarization channel.}
	\label{f:sf}
\end{figure}

\color{black}

\begin{table}
		\setlength{\tabcolsep}{3.6pt}
	\centering
	\caption{
		Fitted models for San Francisco
		and CARABAS~II SAR images
	}
	\label{t:fit}
	\begin{tabular}{cccc| cc }
		\toprule
&	&  $\widehat{\beta}_1$& $\widehat{\beta}_2$
& $\widehat{\beta}^\star_1$& $\widehat{\beta}^\star_2$ \\
\midrule
&	$ 3 \times 3$  & $-2.9162$ &   $0.3091$ &
	$-2.9318$  &  $3.6062$  \\
		\multirow{-2}{*}{SF --- HH $\sim$ HV}
&	$ 7 \times 7$   & $-2.8978$ &   $0.0643$
	& $-2.9115$ &    $1.2344$  \\
		\midrule
&
		$ 3 \times 3$  &$-2.2259$  & $-0.1947$ &
		$-2.1421$ &   $-1.3629$ 	\\
						\multirow{-2}{*}{SF --- VV $\sim$ HV}
&		$ 7 \times 7$   & $-2.1621$ &    $0.0101$ &
		$-2.1559$ &    $0.1569 $   \\
		\midrule
&		$ 3 \times 3$  &$-1.9124$ &  $ -0.4279$ &
		$-1.7851$ &   $ -1.2194$  	\\
			\multirow{-2}{*}{CARABAS~II}
&		$ 7 \times 7$   &  $-1.9896$ &   $0.1616$ &
		$-1.9886$ &    $0.2314$ \\
				\bottomrule
	\end{tabular}
\end{table}

\begin{table}
			\setlength{\tabcolsep}{2.8pt}
	\centering
	\caption{
		Estimated
		root mean square error
		measure
		of the
		fitted models
		for San Francisco
		and CARABAS~II SAR images.
		Best results are
		highlighted
	}
	\label{t:fit2}
\begin{tabular}{cccccc }
	\toprule
	&& Firth-based & MLE-based & & \\
	&	&  Rayleigh & Rayleigh & Gaussian & Rayleigh \\
 && regression & regression & regression & distribution \\
	\midrule
	&	$ 3 \times 3$  & $\mathbf{0.0123}$ &    $0.0133$ &    $1.0019$ &
	$0.5662$  \\
	\multirow{-2}{*}{SF --- HH $\sim$ HV}
	&	$ 7 \times 7$   & $\mathbf{0.0157}$ &    $0.0159$ &    $1.0020$ &
	$0.5606$ \\
	\midrule
	&
	$ 3 \times 3$  	& $\mathbf{0.0194}$ &    $0.0222$ &   $ 1.0077$ &
	$ 0.5085$\\
	\multirow{-2}{*}{SF --- VV $\sim$ HV}
	&		$ 7 \times 7$ &  $\mathbf{0.0326}$ &    $0.0330$ &    $1.0088$
	&   $0.4955$    \\
	\midrule
	&		$ 3 \times 3$  &  $\mathbf{0.0267}$ &   $ 0.0296$ &   $ 1.0136$
	&   $ 0.4730$ 	\\
	\multirow{-2}{*}{CARABAS~II}
	&		$ 7 \times 7$  & $\mathbf{0.0427}$  &  $0.0430$ &   $ 1.0130 $ &
	$ 0.4774$  \\
	\bottomrule
	\end{tabular}
\end{table}

\section{Conclusions}
\label{s:conclu}

This letter introduced
bias-adjusted estimators
for the Rayleigh regression model parameters.
We employed
the Cox and Snell's,
Firth's, and parametric bootstrap methods
to obtain corrected estimators.
Monte Carlo simulation results show that
the discussed bias-adjusted
estimators
outperformed MLE
in terms
of
relative bias and
root mean square
error;
being
the estimators
based on
the
Firth's method
the best performing
approach.
Additionally,
the Firth-based estimators
resulted in more accurate
modeling results
in the evaluated SAR images experiments
than the ones
presented by
the MLE method,
the Rayleigh distribution,
and the Gaussian-based regression model.
In conclusion,
we recommend
the use of corrected estimators based on the
Firth's method
to fit the
Rayleigh regression model
for small signal lengths.

\section*{Appendix}

In this appendix,
we present
the cumulants
of second and third order
used to derive the
Cox and Snell's
and Firth's
corrected estimators.
From~\cite{Palm2019},
we have that
$
\frac{\operatorname{ d } \ell_n (\mu[n])}{\operatorname{ d } \mu[n]}
= \frac{\pi y[n]^2}{2\mu[n]^3}-\frac{2}{\mu[n]}
$,
$
\frac{\operatorname{ d } \mu[n]}{ \operatorname{ d } \eta[n]}
= \frac{1}{g'(\mu[n])}
$,
$
\frac{\partial \eta[n]}{\partial \beta_i}
= x_{i}[n]
$,
$
\frac{\partial^2 \ell_n (\mu[n])}{\partial \mu[n]^2}
= \frac{2}{\mu[n]^2} - \frac{3 \pi y[n]^2}{2 \mu[n]^4}
$,
and
$
\E\left[\frac{\operatorname{ d }^2 \ell_n (\mu[n])}
{\operatorname{ d } \mu[n]^2} \right]
= -\frac{4}{\mu[n]^2}
$.
Note that
$
\frac{\partial^3 \ell_n (\mu[n])}{\partial \mu [n] ^3}
=
\frac{\operatorname{ d }}{\operatorname{ d } \mu [n]}
\left(
\frac{\partial ^2 \ell_n (\mu[n])}{\partial \mu[n]^2}
\right)
=
\frac{6 \pi y[n]^2}{\mu[n]^5}
-
\frac{4}{\mu[n]^3}
$.
Taking the expected value
of the derivative above,
we have
$
\E
\left(
\frac{\partial ^3 \ell_n (\mu[n])}{\partial \mu [n] ^3}
\right)
=
\frac{24}{\mu [n] ^3}
-
\frac{4}{\mu[n]^3}
=
\frac{20}{\mu [n] ^3}
$.
From~\cite{Palm2019},
the second order cumulant
is given by
$
\kappa_ {rs}
=
\E
\left[
\frac{\partial ^2 \ell (\bm{\beta})}
{\partial \beta_r \partial \beta_s}
\right]
=
 \sum_{n=1}^{N}
\left[
-\frac{4}{\mu[n]^2}
\left(
 \dfrac{\operatorname{ d }\mu[n]}{\operatorname{ d }\eta[n]}
\right)^2
x_{s}[n]x_{r}[n]
\right]
$.
Differentiating the
second order cumulant
with respect~$\beta_u$,
we
obtain
$\kappa_{rs}^{(u)}$
as
$
\frac{\partial \kappa_ {rs} }{\partial \beta_u}
 =
 \!\!
\sum \limits _{n=1} ^ N
\!
\left[
\!
\frac{8}{\mu[n]^3}
\!
\left(
\!
\frac{\operatorname{ d } \mu [n]}{\operatorname{ d } \eta [n]}
\!
\right)^3
\!
-
\!
\frac{8}{\mu[n]^2}
\!
\left(
\!
\frac{\operatorname{ d } \mu [n]}{\operatorname{ d } \eta [n]}
\!
\right) ^2
\!
\frac{\partial}{\partial \mu [n]}
\!
\left(
\!
\frac{\operatorname{ d } \mu [n]}{\operatorname{ d } \eta [n]}
\!
\right)
\!
\right]
\!
x_u [n]
x_s [n]
x_r [n]
$.
The third order derivatives
of the log-likelihood function
is
\begin{align*}
&\frac{\partial^3 \ell (\bm{\beta})}
{\partial \beta_r \partial \beta_s
\partial \beta_u}
=
\sum \limits_{n=1} ^ N
\left[
\frac{\partial ^3 \ell_n (\mu[n])}{\partial \mu [n] ^3}
\left(
\frac{\operatorname{ d } \mu [n]}{\operatorname{ d } \eta [n]}
\right)^2
+
\frac{\partial}{\partial \mu [n]}
\right.
\\
&\left.
\cdot
\left(
\frac{\operatorname{ d } \mu [n]}{\operatorname{ d } \eta [n]}
\right)^2
\frac{\partial ^2 \ell_n (\mu[n])}{\partial \mu [n] ^2}
+
\frac{\partial ^2 \ell_n (\mu[n])}{\partial \mu [n] ^2}
\right.
\left.
\frac{\operatorname{ d } \mu [n]}{\operatorname{ d } \eta[n]}
\frac{\partial}{\partial \mu [n]}
\left(
\frac{\operatorname{ d } \mu [n] }{\operatorname{ d } \eta [n]}
\right)
\right.
\\
&\left.
+
\frac{\partial}{\partial \mu [n]}
\!\left(
\!\!\frac{\operatorname{ d } \mu [n]}{\operatorname{ d } \eta [n]}
\frac{\partial}{\partial \mu [n]}
\frac{\operatorname{ d } \mu [n]}{\operatorname{ d } \eta [n]}
\right)
\!\!\frac{\operatorname{ d } \ell_n (\mu[n]) }{\operatorname{ d } \mu [n]}
\right]
\!\!\frac{\operatorname{ d } \mu [n]}{\operatorname{ d } \eta [n]}
x_u [n]
x_r [n]
x_s [n]
.
\end{align*}
Taking the expected value,
we obtain the third order
cumulant
\[
\kappa_{rsu}
=
\sum \limits_{n=1} ^ N
\!
\left[
\!
\frac{20} {\mu [n] ^3}
\!
\left(
\!
\frac{\operatorname{ d } \mu [n]}{\operatorname{ d } \eta [n]}
\!
\right)^3
\!\!\!
-
\!
\frac{12}{\mu [n] ^2}
\!
\left(
\!
\frac{\operatorname{ d } \mu [n]}{\operatorname{ d } \eta [n]}
\!
\right)^2
\!
\frac{\partial}{\partial \mu [n]}
\!
\left(
\!
\frac{\operatorname{ d } \mu [n]}{\operatorname{ d } \eta [n]}
\!
\right)
\!
\right]
\!
x_u [n]
x_r [n]
x_s [n]
.
\]
From the above expressions,
we have that
$
\kappa_{rs}^{(u)}
-
\frac{1}{2}
\kappa_{rsu}
=
\sum \limits_{n=1} ^N
\!
\left[
\!
-
\frac{2}{\mu[n]^3}
\left(
\frac{\operatorname{ d } \mu [n]}{\operatorname{ d } \eta [n]}
\right)^3
\!\!\!\!
-
\!
\frac{2}{\mu [n] ^2}
\!
\left(
\!
\frac{\operatorname{ d } \mu [n]}{\operatorname{ d } \eta [n]}
\!
\right)^2
\!\!\!\!
\frac{\partial}{\partial \mu [n]}
\!\!
\left(
\!
\frac{\operatorname{ d } \mu [n]}{\operatorname{ d } \eta [n]}
\!
\right)
\!
\right]
\!\!
x_u [n]
x_s [n]
x_r [n]
$.
Now, it is possible to compute
the second order
biases of
the Rayleigh regression model~MLE
as
$
\sum \limits _{r,s,u}
\kappa ^{ar}
\kappa ^{su}
\left\lbrace
\kappa_ {rs} ^ {(u)}
-
\frac{1}{2}
\kappa_{rsu}
\right\rbrace
=
\sum \limits_{n=1} ^N
w [n]
\sum \limits_r
\kappa^{ar}
x_r [n]
\sum \limits _{s,u}
x_s [n]
\kappa^{su}
x_u [n]
$,
where
$
w[n]
=
-
\frac{2}{\mu[n]^3}
\left(
\frac{\operatorname{ d } \mu [n]}{\operatorname{ d } \eta [n]}
\right)^3
-
\frac{2}{\mu [n] ^2}
\left(
\frac{\operatorname{ d } \mu [n]}{\operatorname{ d } \eta [n]}
\right)^2
\frac{\partial}{\partial \mu [n]}
\left(
\frac{\operatorname{ d } \mu [n]}{\operatorname{ d } \eta [n]}
\right)
$.
Note that
$
\sum \limits_{n=1} ^N
w [n]
\sum \limits_r
\kappa^{ar}
x_r [n]
\sum \limits _{s,u}
x_s [n]
\kappa^{su}
x_u [n]
=
e_a^\top
\mathbf{I}^{-1} (\bm{\beta})
\sum \limits_{n=1} ^N
w [n]
x [n]
\left(
x^\top [n]
\mathbf{I}^{-1} (\bm{\beta})
x [n]
\right)
$,
where~$e_a$
is defined as the~$a$th column vector
of the~$k \times k$ identity matrix.
Then,
$
\sum \limits _{r,s,u}
\kappa ^{ar}
\kappa ^{su}
\left\lbrace
\kappa_ {rs} ^ {(u)}
-
\frac{1}{2}
\kappa_{rsu}
\right\rbrace
=
e_a^\top
\cdot
\mathbf{I}^{-1} (\bm{\beta})
\cdot
\mathbf{X}^\top
\cdot
\mathbf{W}
\cdot
\delta
$.

{\small
\singlespacing
\bibliographystyle{siam}
\bibliography{rayleigh}

\begin{thebibliography}{10}

\bibitem{dimas2019}
{\sc D.~I. Alves, B.~G. Palm, M.~I. Pettersson, V.~T. Vu, R.~Machado, B.~F.
  Uchoa-Filho, P.~Dammert, and H.~Hellsten}, {\em A statistical analysis for
  wavelength-resolution {SAR} image stacks}, IEEE Geoscience and Remote Sensing
  Letters, 17 (2019), pp.~227--231.

\bibitem{Cintra2013}
{\sc R.~J. Cintra, A.~C. Frery, and A.~D. Nascimento}, {\em Parametric and
  nonparametric tests for speckled imagery}, Pattern Analysis and Applications,
  16 (2013), pp.~141--161.

\bibitem{CordeiroCribari2014}
{\sc G.~M. Cordeiro and F.~Cribari-Neto}, {\em An Introduction to Bartlett
  Correction and Bias Reduction}, Springer, 2014.

\bibitem{cox1968}
{\sc D.~R. Cox and E.~J. Snell}, {\em A general definition of residuals},
  Journal of the Royal Statistical Society: Series B (Methodological), 30
  (1968), pp.~248--265.

\bibitem{cribari2002}
{\sc F.~Cribari-Neto and K.~L. Vasconcellos}, {\em Nearly unbiased maximum
  likelihood estimation for the beta distribution}, Journal of Statistical
  Computation and Simulation, 72 (2002), pp.~107--118.

\bibitem{Efron1979}
{\sc B.~Efron}, {\em Bootstrap methods: Another look at the jackknife}, The
  Annals of Statistics, 7 (1979), pp.~1--26.

\bibitem{Efron1994}
{\sc B.~Efron and R.~J. Tibshirani}, {\em An Introduction to the {B}ootstrap},
  Monographs on Statistics and Applied Probability 57, 1994.

\bibitem{Firth1993}
{\sc D.~Firth}, {\em Bias reduction of maximum likelihood estimates},
  Biometrika, 80 (1993), pp.~27--38.

\bibitem{gomes2018}
{\sc N.~R. Gomes, P.~Dammert, M.~I. Pettersson, V.~T. Vu, and H.~Hellsten},
  {\em Comparison of the {R}ayleigh and $k$-distributions for application in
  incoherent change detection}, IEEE Geoscience and Remote Sensing Letters, 16
  (2018), pp.~756--760.

\bibitem{li2012}
{\sc G.-T. Li, C.-L. Wang, P.-P. Huang, and W.-D. Yu}, {\em {SAR} image
  despeckling using a space-domain filter with alterable window}, IEEE
  Geoscience and Remote Sensing Letters, 10 (2012), pp.~263--267.

\bibitem{Lundberg2006}
{\sc M.~Lundberg, L.~M.~H. Ulander, W.~E. Pierson, and A.~Gustavsson}, {\em A
  challenge problem for detection of targets in foliage}, in Proc. SPIE,
  vol.~6237, 2006.

\bibitem{oliver2004}
{\sc C.~Oliver and S.~Quegan}, {\em Understanding synthetic aperture radar
  images}, SciTech Publishing, 2004.

\bibitem{Palm2019}
{\sc B.~G. Palm, F.~M. Bayer, R.~J. Cintra, M.~I. Pettersson, and R.~Machado},
  {\em Rayleigh regression model for ground type detection in {SAR} imagery},
  IEEE Geoscience and Remote Sensing Letters, 16 (2019), pp.~1660--1664.

\bibitem{park1999}
{\sc J.-M. Park, W.-J. Song, and W.~Pearlman}, {\em Speckle filtering of {SAR}
  images based on adaptive windowing}, IEE Proceedings--Vision, Image and
  Signal Processing, 146 (1999), pp.~191--197.

\bibitem{sumaiya2018}
{\sc M.~N. Sumaiya and R.~S.~S. Kumari}, {\em Unsupervised change detection of
  flood affected areas in {SAR} images using {R}ayleigh-based {B}ayesian
  thresholding}, IET Radar, Sonar \& Navigation, 12 (2018), pp.~515--522.

\bibitem{wiesel2008}
{\sc A.~Wiesel, Y.~C. Eldar, and A.~Yeredor}, {\em Linear regression with
  {G}aussian model uncertainty: {A}lgorithms and bounds}, IEEE Transactions on
  Signal Processing, 56 (2008), pp.~2194--2205.

\bibitem{xiang2016built}
{\sc D.~Xiang, T.~Tang, C.~Hu, Q.~Fan, and Y.~Su}, {\em Built-up area
  extraction from {P}ol{SAR} imagery with model-based decomposition and
  polarimetric coherence}, Remote Sensing, 8 (2016), p.~685.

\end{thebibliography}
}

\end{document}